\documentclass[12pt]{article}
\usepackage{graphicx}
\usepackage{amssymb}
\usepackage{amsmath}
\usepackage{bm}
\usepackage{cite}

\newcommand{\bear}{\begin{array}}  
\newcommand {\eear}{\end{array}}
\newcommand{\bea}{\begin{eqnarray}}   
\newcommand{\eea}{\end{eqnarray}}
\newcommand{\beq}{\begin{eqnarray}}   
\newcommand{\eeq}{\end{eqnarray}}
\newcommand{\bef}{\begin{figure}}  \newcommand 
{\eef}{\end{figure}}
\newcommand{\bec}{\begin{center}}  \newcommand 
{\eec}{\end{center}}

\begin{document}

\begin{titlepage}

\begin{flushright}
IPMU 13-0075
\end{flushright}

\vskip 1.35cm
\begin{center}

{\large 
{\bf Grand Unification in High-scale Supersymmetry}
}

\vskip 1.2cm

Junji Hisano$^{a,b}$, 
Takumi Kuwahara$^a$,
and
Natsumi Nagata$^{a,c}$\\

\vskip 0.4cm
{\it $^a$Department of Physics,
Nagoya University, Nagoya 464-8602, Japan}\\
{\it $^b$Kavli Institute for the Physics and Mathematics of the Universe (WPI),
Todai Institutes for Advanced Study, the University of Tokyo}, Kashiwa 277-8568, Japan\\
{\it $^c$Department of Physics, 
University of Tokyo, Tokyo 113-0033, Japan}

\date{\today}

\vskip 1.5cm

\begin{abstract} 

 A constraint on masses of superheavy gauge and Higgs multiplets at
 the grand unification (GUT) scale is obtained from the gauge coupling
 unification in the case of high-scale supersymmetry. We found that all
 of the particles may lie around a scale of $10^{16}$~GeV so that the
 threshold corrections to the gauge coupling constants at the GUT scale
 are smaller than those in the case of the low-energy supersymmetry. In
 addition, the GUT scale tends to be slightly lower when the gauginos 
 are heavier and, thus, the proton decay rate via the $X$-boson exchange
 process is expected to be enhanced.  
\end{abstract}

\end{center}
\end{titlepage}

%%%%%%%%%%%%%%%%%%%%%%%%%%%%%%%%%%%
\section{Introduction}
%%%%%%%%%%%%%%%%%%%%%%%%%%%%%%%%%%%

Supersymmetric grand unified theories (SUSY GUTs)
\cite{Dimopoulos:1981zb, Sakai:1981gr} are promising candidates of 
physics beyond the Standard Model (SM). Indeed, they are strongly
motivated by an experimental observation which implies that the gauge
coupling constants of the SM gauge groups are to be unified at a certain
high-energy scale with good accuracy \cite{Dimopoulos:1981yj,
Marciano:1981un, Einhorn:1981sx, Amaldi:1991cn, Langacker:1991an}. 
From the observation, the unified scale is estimated as $M_{\rm GUT}
\simeq 2\times 10^{16}$~GeV. In the GUTs, new superheavy particles are
supposed to appear around the scale, which make the couplings run
together above the scale. These particles are naturally expected to have
masses of the order of $M_{\rm GUT}$. They are, of course, beyond the
reach of collider experiments, and there is little hope to search for them or
to measure their masses directly.

In Refs.~\cite{Hisano:1992mh,Hisano:1992jj}, a way of constraining the
masses of superheavy particles indirectly by requiring the gauge
coupling unification is discussed and limits on the masses are
presented. Later, by applying the same method with more accurate gauge 
coupling constants, the
authors in Ref.~\cite{Murayama:2001ur} derive a more stringent
constraint on the masses of the color-triplet Higgs boson, the adjoint
Higgs bosons, and the $X$ bosons in the
context of the minimal SUSY SU(5) GUT \cite{Dimopoulos:1981zb,
Sakai:1981gr}. They have found that while the masses of the adjoint Higgs
and the $X$ bosons are around $10^{16}$~GeV, the mass of the
color-triplet Higgs boson lies in the region of $3.5\times
10^{14}~{\rm GeV} \lesssim M_{H_C}\lesssim 3.6 \times 10^{15}$~GeV, which is
significantly below the GUT scale. This implies that the threshold
corrections to the gauge coupling constants are still not small, even in
the SUSY GUTs. In fact, the analysis is quite
sensitive to the mass spectrum in the intermediate scale, such as those
of the SUSY particles. In Ref.~\cite{Murayama:2001ur}, all of the
particles except for gauginos are assumed to be at 1~TeV. The gaugino
masses are set to be around the electroweak scale with the GUT relation
for the gaugino masses being assumed.

Currently, on the other hand, the SUSY models with heavy sfermions have
been widely discussed with a lot of attention \cite{Giudice:1998xp,
Wells:2003tf, ArkaniHamed:2004fb, Giudice:2004tc, ArkaniHamed:2004yi,
Wells:2004di, Hall:2009nd, Hall:2011jd}. 
Although such models originally have been
discussed as a possible candidate of SUSY models from a theoretical point
of view, now they are also supported by the latest results of the LHC
experiments; no significant excess in the SUSY searches 
\cite{:2012rz, ATLAS:2012ona, :2012mfa} and the discovery of the 126~GeV
Higgs boson \cite{:2012gk,:2012gu} both suggest that the SUSY breaking
scale is somewhat higher than the electroweak scale. While this
high-scale SUSY scenario is difficult to be probed at the collider
experiments, it still has a lot of phenomenological consequences which
might be checked in other experiments. Recent studies on the subject are
given in Refs.~\cite{Giudice:2011cg, Ibe:2011aa, Ibe:2012hu,
Ibanez:2013gf,Hisano:2012wm,Jeong:2011sg, Saito:2012bb, Sato:2012xf,
Bhattacherjee:2012ed, Hall:2012zp, ArkaniHamed:2012gw,
Moroi:2013sfa,McKeen:2013dma}.

As mentioned above, the GUT scale mass spectrum inferred from the
indirect analysis presented in Refs.~\cite{Hisano:1992mh,Hisano:1992jj}
is highly dependent on the SUSY mass spectrum. In particular, if the
SUSY particles have masses much larger than ${\cal O}(1)$~TeV, previous
results are expected to be changed significantly. In this Letter,
therefore, we revisit the analysis in the high-scale SUSY scenario. We
carry out the calculation in the minimal SUSY SU(5) GUT with sfermions
having masses well above the electroweak scale. We will see that while
the constraint on the adjoint Higgs and $X$ boson masses only differs
from previous ones slightly, that on the color-triplet Higgs mass
is found to be changed by more than an order of magnitude and actually
improved in the sense that it may also be around the GUT
scale. Interestingly enough, this result again indicates that the
high-scale SUSY scenario is rather supported than the traditional
low-energy SUSY models. 

This Letter is organized as follows: in Sec.~\ref{model}, the high-scale
SUSY model which we discuss below is presented, and its phenomenological
aspects are briefly described. The mass spectrum of the superheavy
particles in the minimal SUSY SU(5) GUT is also displayed there. In the
subsequent section, we discuss the method of constraining the masses of
the GUT-scale multiplets by means of the renormalization group equations
(RGEs) as well as the threshold corrections of the gauge
couplings. Then, we show some results in
Sec.~\ref{results}. Section.~\ref{conclusion} is devoted to conclusions
and discussion.

%%%%%%%%%%%%%%%%%%%%%%%%%%%%%%%%%%%
\section{Model and Spectrum}
\label{model}
%%%%%%%%%%%%%%%%%%%%%%%%%%%%%%%

Let us begin by presenting a high-scale SUSY model discussed in this
Letter. We consider the particle content of the minimal supersymmetric
Standard Model (MSSM). Then, the only assumption which we adopt here is
that there exists a dynamical SUSY-breaking sector with the K\"{a}hler
potential having a generic structure. Then, all of the scalar bosons except
the lightest Higgs boson acquire masses of the order of the gravitino
mass $m_{3/2}$, while the lightest Higgs boson mass is fine-tuned to be
$m_h\sim 126$~GeV. The higgsinos may in general have similar masses to
the gravitino mass, though they might be suppressed if there are some
extra chiral symmetries. The gaugino masses are, on the other hand,
generated by the quantum effects \cite{Randall:1998uk, Giudice:1998xp}
and, thus,  suppressed by a loop factor compared with $m_{3/2}$. There
are two kinds of contributions to the gaugino masses; one is the anomaly
mediation effect \cite{Randall:1998uk, Giudice:1998xp} and the other is
from the higgsino-Higgs boson loop diagram. These two effects give rise
to the gaugino masses as
\begin{align}
 M_1&=\frac{33}{5}\frac{g_1^2}{16\pi^2}\biggl(
m_{3/2}+\frac{1}{11}L
\biggr)~,\nonumber \\
 M_2&=\frac{g_2^2}{16\pi^2}(
m_{3/2}+L)~,\nonumber \\ 
M_3&=-3\frac{g_3^2}{16\pi^2}
m_{3/2}~,
\label{gaugino_masses}
\end{align}
with
\begin{equation}
 L=\mu_H \sin 2\beta\frac{m_A^2}{\vert \mu_H\vert^2-m_A^2}
\ln\frac{\vert \mu_H\vert^2}{m_A^2}~
\end{equation}
representing the higgsino-Higgs boson loop contribution. 
Here, $M_a$
and $g_a$ ($a=1,2,$ and 3) are the U(1)$_Y$, SU(2)$_L$, and SU(3)$_C$
gaugino masses and gauge coupling constants, respectively. We use the
SU(5) normalization for the U(1)$_Y$ coupling, {\it i.e.}, $g_1\equiv
\sqrt{5/3}g^\prime$. Further, $\mu_H$ and $m_A$ denote the higgsino
and the heavy Higgs boson masses, respectively, which are assumed to
be the same order of magnitude as sfermion masses. $\tan \beta$ is the
ratio of the vacuum expectation values (VEVs) of the Higgs fields;
$\tan\beta\equiv \langle H_u\rangle /\langle H_d \rangle$. Since the
higgsino mass is presumed to be around the gravitino mass, the
higgsino-loop contribution $L$ is also expected to be as large as
$m_{3/2}$. The values of the gaugino masses are, however, dependent on
the relative phase between $\mu_H$ and $m_{3/2}$. In the following
discussion we just assume the gaugino masses are lighter than the
scalar masses by a loop factor, and regard them as free parameters.

As mentioned in Introduction, this model has a lot of
fascinating features from a phenomenological point of view. They are
originally discussed in Refs.~\cite{Giudice:1998xp, Wells:2003tf,
ArkaniHamed:2004fb, Giudice:2004tc, ArkaniHamed:2004yi, Wells:2004di,
Hall:2009nd, Hall:2011jd}, and recent development is given in
Refs.~\cite{Giudice:2011cg, Ibe:2011aa, Ibe:2012hu, 
Ibanez:2013gf,Hisano:2012wm,Jeong:2011sg, Saito:2012bb, Sato:2012xf,
Bhattacherjee:2012ed, Hall:2012zp, ArkaniHamed:2012gw,
Moroi:2013sfa,McKeen:2013dma}. In these works, a typical scale for
sfermion masses is taken to be ${\cal O}(10^2$--$10^3$)~TeV, which
explains the 126~GeV Higgs boson mass. With such heavy particles the SUSY
flavor and CP problems \cite{Gabbiani:1996hi} as well as the gravitino
problem are considerably relaxed. In this case, the gaugino masses are
${\cal O}$(1) TeV because of an one-loop factor. Note that from
Eq.~\eqref{gaugino_masses} it is found that with a moderate value of
$L$, wino becomes the lightest among gauginos, and thus the lightest
SUSY particle in this model. It is quite interesting since the wino dark
matter with a mass of $2.7$--3~TeV is consistent with cosmological
observations \cite{Hisano:2006nn}. The dark matter might be searched
directly \cite{Hisano:2012wm, Hisano:2010fy, Hisano:2010ct} or
indirectly \cite{Hisano:2003ec, Hisano:2004ds, Ibe:2012hu} in future
dark matter experiments.

Moreover, the gauge coupling unification in this high-scale SUSY model
is found to be achieved as precisely as that in the MSSM. Thus
the SUSY GUT is still promising in the case of high-scale
supersymmetry. In the next section, we in turn require the gauge
coupling unification and constrain the GUT scale mass spectrum by using
the requirement.

Now, we briefly summarize the superheavy particles in the minimal SUSY
SU(5) GUT, which we adopt as the working hypothesis in this paper. The
SUSY SU(5) gauge theory includes twenty-four gauge superfields ${\cal
  V}^A$ with $A=1,\dots, 24$. By using the SU(5) generators $T^A$ we
define a $5\times 5$ matrix representation of the vector superfields
such that ${\cal V}\equiv {\cal V}^AT^A$, with the components written
as
\begin{equation}
 {\cal V}=\frac{1}{\sqrt{2}}
\begin{pmatrix}
 \begin{matrix}
G  -\frac{2}{\sqrt{30}} B
 \end{matrix}
&
\begin{matrix}
X^{\dagger 1} \\
X^{\dagger 2} \\
X^{\dagger 3}
\end{matrix}
&
\begin{matrix}
 Y^{\dagger 1}\\ Y^{\dagger 2} \\ Y^{\dagger 3}
\end{matrix}
\\
\begin{matrix}
 X_1 & X_2 & X_3 \\
 Y_1 & Y_2 & Y_3  
\end{matrix}
&
\begin{matrix}
 \frac{1}{\sqrt{2}}W^3+\frac{3}{\sqrt{30}}B \\ W^-
\end{matrix}
&
\begin{matrix}
W^+ \\ - \frac{1}{\sqrt{2}}W^3+\frac{3}{\sqrt{30}}B
\end{matrix}
\end{pmatrix}
~.
\end{equation} 
We collectively call $X_\alpha$ and $Y_\alpha$ the $X$-bosons hereafter.
The unified gauge group SU(5) is spontaneously broken by the VEV of the
adjoint Higgs boson $\Sigma^A$ ($A=1,\dots,24$) to
SU(3)$_C\times$SU(2)$_{L}\times$U(1)$_Y$. Again, we define $\Sigma\equiv
\Sigma^A T^A$ and its components by
\begin{equation}
\Sigma =
\begin{pmatrix}
 \Sigma_8&\Sigma_{(3,2)} \\
 \Sigma_{(3^*,2)} & \Sigma_3
\end{pmatrix}
+\frac{1}{2\sqrt{15}}
\begin{pmatrix}
 2&0\\0&-3
\end{pmatrix}
\Sigma_{24}~.
\end{equation}
The MSSM Higgs superfields are, on the other hand, incorporated into
fundamental and  anti-fundamental representations as follows:
\begin{equation}
 H=
\begin{pmatrix}
 H^1_C \\ H^2_C \\ H^3_C \\ H^+_u \\ H^0_u
\end{pmatrix}
,~~~~~~\bar{H}=
\begin{pmatrix}
 \bar{H}_{C1}\\
 \bar{H}_{C2}\\
 \bar{H}_{C3}\\
 H^-_d \\ -H^0_d
\end{pmatrix}
~,
\end{equation}
where
\begin{equation}
 H_u=
\begin{pmatrix}
 H^+_u \\ H^0_u
\end{pmatrix}
,~~~~~~H_d=
\begin{pmatrix}
 H^0_d \\ H^-_d
\end{pmatrix}
~,
\end{equation}
are the MSSM Higgs superfields. $H^\alpha_C$ and $\bar{H}_{C\alpha}$ are
called the color-triplet Higgs multiplets.

The superpotential of the Higgs sector in the minimal SU(5) SUSY GUT is
given by
\begin{align}
 W &= \frac{1}{3}\lambda_{\Sigma}{\rm Tr}\Sigma^3 +\frac{1}{2}m_\Sigma
 {\rm Tr} \Sigma^2 +\lambda_H \bar{H}\Sigma H +m_H\bar{H} H~.
\label{superpotential}
\end{align}
After the adjoint Higgs field gets the VEV, $ \langle \Sigma \rangle
=V\cdot {\rm diag}(2,2,2,-3,-3)$ ($V=m_\Sigma/\lambda_\Sigma$), the
SU(5) gauge group is broken to the SM 
SU(3)$_C\times$SU(2)$_{L}\times$U(1)$_Y$ gauge groups without breaking
the supersymmetry. Also the parameter $m_H$ is fine-tuned as $m_H
=3\lambda_H V$ in order to realize  the doublet-triplet mass splitting
in $H$ and $\bar{H}$. With
the condition, the mass of the color-triplet Higgs boson is
$M_{H_C}=M_{\bar{H}_C}=5\lambda_H V$.  The $X$-boson mass is
$M_X=5\sqrt{2}g_5 V$ with $g_5$ the unified gauge coupling
constant. As regards the adjoint Higgs multiplets, the components $\Sigma_3$
and $\Sigma_8$ have masses of $M_{\Sigma}\equiv
M_{\Sigma_8}=M_{\Sigma_3}=\frac{5}{2}\lambda_{\Sigma}V$ and
$\Sigma_{24}$ has $M_{\Sigma_{24}}=\frac{1}{2}\lambda_{\Sigma}V$. The
components $\Sigma_{(3^*, 2)}$ and $\Sigma_{(3,2)}$ become the 
longitudinal component of the $X$-bosons, and thus do not show up as
physical states.

%%%%%%%%%%%%%%%%%%%%%%%%%%%%%%%%%%%%%%%%%%%
\section{Renormalization Group Analysis}
%%%%%%%%%%%%%%%%%%%%%%%%%%%%%%%%%%%%%%%%%%%

In this section, we present the 
RGEs of the gauge and Yukawa coupling constants, as well as the
boundary conditions at each threshold. We use the $\overline{\rm DR}$
scheme \cite{Siegel:1979wq} in this work.
First, we write down the two-loop beta
functions \cite{Machacek:1983tz}. 
In the MSSM, the two-loop RGEs for the gauge coupling
constants are given as 
\begin{equation}
 \mu \frac{\partial g_a}{\partial \mu}=\frac{1}{16\pi^2}b_a 
^{(1)}g^3_a
+\frac{g_a^3}{(16\pi^2)^2}\biggl[
\sum_{b=1}^{3}b_{ab}^{(2)}g_b^2 -\sum_{i=t,b,\tau}c_{ai}~ y^2_i 
\biggr]~,
\end{equation}
where
\begin{equation}
b^{(1)}_a=
\begin{pmatrix}
 33/5 \\ 1 \\ -3
\end{pmatrix}
~,~~~~b_{ab}^{(2)}=
\begin{pmatrix}
 199/25 & 27/5 & 88/5 \\
 9/5 & 25 & 24 \\
 11/5 & 9 & 14 
\end{pmatrix} 
~,
\label{MSSM_beta}
\end{equation}
and
\begin{equation}
 c_{ai}=
\begin{pmatrix}
 26/5 & 14/5 & 18/5 \\
 6 & 6 & 2 \\
 4 & 4 & 0
\end{pmatrix}
~, ~~~~~~
\end{equation}
with $y_i$ ($i=t,b,\tau$) the top, bottom and tau Yukawa coupling
constants, respectively. Since the Yukawa 
couplings enter into the two-loop level contributions to the gauge
coupling RGEs, it is sufficient to consider the RGEs for the Yukawa
couplings at one-loop level. They are given as 
\begin{align}
 \mu\frac{\partial}{\partial \mu}y_t&=\frac{1}{16\pi^2}y_t
\biggl[6y_t^2+y^2_b -\frac{13}{15}g^2_1-3g_2^2-\frac{16}{3}
 g_3^2\biggr] , \nonumber \\
 \mu\frac{\partial}{\partial \mu}y_b &=\frac{1}{16\pi^2}y_b
\biggl[6y_b^2+y_t^2+y_\tau^2
 -\frac{7}{15}g^2_1-3g_2^2-\frac{16}{3}
 g_3^2\biggr] , \nonumber \\
 \mu\frac{\partial}{\partial \mu}y_\tau &=\frac{1}{16\pi^2}y_\tau
\biggl[3y_b^2+4y_\tau^2 
 -\frac{9}{5}g^2_1-3g_2^2\biggr] .
\end{align}
Below the SUSY breaking scale ($M_S$), the squarks and sleptons, the
higgsinos, and the heavy Higgs boson masses are decoupled so that the
theory is regarded as the SM with gauginos. The contribution of gauginos
and the SM particles to the coefficients of the beta functions is given as
\begin{equation}
 b_a^{(1)}=
\begin{pmatrix}
 41/10 \\ -19/6 \\ -7
\end{pmatrix}
_{\rm SM}
+
\begin{pmatrix}
 0 \\ 4/3 \\ 2
\end{pmatrix}
_{\rm gaugino}
~,
\end{equation}
\begin{align}
 b^{(2)}_{ab}&=
\begin{pmatrix}
 199/50 & 27/10 & 44/5 \\
 9/10 & 35/6 & 12 \\
 11/10 & 9/2 & -26
\end{pmatrix}
_{\rm SM}
+
\begin{pmatrix}
 0 & 0 & 0 \\
 0 & 64/3 & 0 \\
 0 & 0 & 48
\end{pmatrix}
_{\rm gaugino}
~,
\label{SMbeta2}
\end{align}
and
\begin{equation}
 c_{ai}=
\begin{pmatrix}
 17/10 & 1/2 & 3/2 \\
 3/2 & 3/2 & 1/2 \\
 2 & 2 & 0
\end{pmatrix}
_{\rm SM}
~,
\end{equation}
where the subscripts ``SM'' and  ``gaugino'' indicate that the
contributions are of the SM particles and gauginos, 
respectively. 
The running of the Yukawa couplings in this case is given as follows:
\begin{align}
 \mu\frac{\partial}{\partial \mu}y_t&=\frac{1}{16\pi^2}y_t
\biggl[
\biggl(
\frac{9}{2}y_t^2 +\frac{3}{2}y_b^2 +y_\tau^2
 -\frac{17}{20}g^2_1-\frac{9}{4}g_2^2-8
 g_3^2\biggr)_{\rm SM}
~
\biggr] , \nonumber \\
 \mu\frac{\partial}{\partial \mu}y_b&=\frac{1}{16\pi^2}y_b
\biggl[
\biggl(
\frac{3}{2}y_t^2+\frac{9}{2}y_b^2 +y_\tau^2
 -\frac{1}{4}g^2_1-\frac{9}{4}g_2^2-8
 g_3^2\biggr)_{\rm SM}
~
\biggr] , \nonumber \\ 
 \mu\frac{\partial}{\partial \mu}y_\tau&=\frac{1}{16\pi^2}y_\tau
\biggl[
\biggl(3y_t^2 +3y_b^2+\frac{5}{2}y_\tau^2
 -\frac{9}{4}g^2_1-\frac{9}{4}g_2^2\biggr)_{\rm SM}
~
\biggr] .
\end{align}

Next, we consider the matching conditions at each threshold scale. At
the GUT scale, the gauge coupling constants in the
SU(3)$_C\times$SU(2)$_L\times$U(1)$_Y$ gauge theories are equated to the
unified coupling constant $g_5$ with the following threshold corrections
at one-loop level \cite{Weinberg:1980wa, Hall:1980kf}:
\begin{align}
 \frac{1}{g_1^2(\mu)}&=\frac{1}{g_5^2(\mu)}+\frac{1}{8\pi^2}\biggl[
\frac{2}{5}
\ln \frac{\mu}{M_{H_C}}-10\ln\frac{\mu}{M_X}
\biggr]~,\nonumber \\
 \frac{1}{g_2^2(\mu)}&=\frac{1}{g_5^2(\mu)}+\frac{1}{8\pi^2}\biggl[
2\ln \frac{\mu}{M_\Sigma}-6\ln\frac{\mu}{M_X}
\biggr]~,\nonumber \\
 \frac{1}{g_3^2(\mu)}&=\frac{1}{g_5^2(\mu)}+\frac{1}{8\pi^2}\biggl[
\ln \frac{\mu}{M_{H_C}}+3\ln \frac{\mu}{M_\Sigma}-4\ln\frac{\mu}{M_X}
\biggr]~.
\end{align}
Here, the conditions do not include constant (scale independent) terms
since we use the $\overline{\rm DR}$ scheme for the renormalization
\cite{Antoniadis:1982vr, Einhorn:1981sx}. From the equations it
immediately follows that
\begin{align}
 \frac{3}{g_2^2(\mu)}- \frac{2}{g_3^2(\mu)}- \frac{1}{g_1^2(\mu)}
&=-\frac{3}{10\pi^2}\ln \frac{\mu}{M_{H_C}}~, \nonumber \\
 \frac{5}{g_1^2(\mu)}- \frac{3}{g_2^2(\mu)}- \frac{2}{g_3^2(\mu)}
&=-\frac{3}{2\pi^2}\ln \frac{\mu^3}{M_X^2M_{\Sigma}}~.
\label{conditions}
\end{align}
The relations allow us to evaluate the masses of the heavy particles,
$M_{H_C}$ and $M_X^2M_{\Sigma}$, from the gauge coupling constants
determined in the low-energy experiments through the RGEs
\cite{Hisano:1992mh,Hisano:1992jj}. While the couplings are well
measured with high precision, the estimation is quite dependent on the
spectrum in the intermediate region, especially on the masses of
gauginos and higgsinos. 

For the SUSY breaking threshold, we just equate the gauge couplings
above and below the threshold, and change the beta functions
appropriately for each region. This approximation is valid since the
particles appearing at the scale are assumed to be degenerate in mass. 
In the case of gauginos, on the other hand, we need to consider the
threshold corrections since the mass difference among gauginos might be
sizable. The condition is
\begin{align}
 \frac{1}{g_1^2(\mu)_{\rm SM}}&=\frac{1}{g_1^2(\mu)_{\rm gaugino}}
~,\nonumber \\
 \frac{1}{g_2^2(\mu)_{\rm SM}}&=\frac{1}{g_2^2(\mu)_{\rm gaugino}}
+\frac{1}{6\pi^2}\ln \frac{\mu}{M_2}
~,\nonumber \\
 \frac{1}{g_3^2(\mu)_{\rm SM}}&=\frac{1}{g_3^2(\mu)_{\rm gaugino}}
+\frac{1}{4\pi^2}\ln\frac{\mu}{M_3}
~,
\end{align}
where $g_a(\mu)_{\rm SM}$ are the couplings in the SM while
$g_a(\mu)_{\rm gaugino}$ are those above the gaugino threshold.

The Yukawa couplings are matched as usual, {\it i.e.}, at the SUSY breaking
scale,  the Yukawa couplings $y_i(\mu)$ below the SUSY breaking
scale are matched with
the supersymmetric ones, $y_i(\mu)_{\rm MSSM}$, as follows:
\begin{align}
 y_{t}(M_S)_{\rm MSSM}&=\frac{1}{\sin\beta}{y}_{t}(M_S)~, \nonumber \\
 y_{b}(M_S)_{\rm MSSM}&=\frac{1}{\cos\beta}{y}_{b}(M_S)~,  \nonumber \\
 y_{\tau}(M_S)_{\rm MSSM}&=\frac{1}{\cos\beta}{y}_{\tau}(M_S)~.
\end{align}

Before concluding this section, we solve the RGEs at one-loop level and,
taking the threshold corrections into account, derive relations between
the superheavy masses and the low-energy gauge coupling constants. Such
relations reflect the dependence of $M_{H_C}$ and $M_X^2M_{\Sigma}$ on
the mass spectrum of the SUSY particles. By inserting to
Eq.~\eqref{conditions} the one-loop solutions of the RGEs for the gauge
couplings, we have
\begin{align}
 \frac{3}{\alpha_2(m_Z)}-\frac{2}{\alpha_3(m_Z)}-\frac{1}{\alpha_1(m_Z)}&
=\frac{1}{2\pi}\biggl[
\frac{12}{5}\ln\biggl(\frac{M_{H_C}}{m_Z}\biggr)-2\ln\biggl(
\frac{M_S}{m_Z}
\biggr)+4\ln\biggl(\frac{M_3}{M_2}\biggr)
\biggr]~,\label{mhc}\\
 \frac{5}{\alpha_1(m_Z)}-\frac{3}{\alpha_2(m_Z)}-\frac{2}{\alpha_3(m_Z)}&
=\frac{1}{2\pi}\biggl[
12\ln\biggl(\frac{M_X^2M_\Sigma}{m_Z^3}\biggr)+4\ln\biggl(
\frac{M_2}{m_Z}
\biggr)+4\ln\biggl(\frac{M_3}{m_Z}\biggr)
\biggr]~.\label{mgut}
\end{align}
From Eq.~\eqref{mhc} we find that the mass of the color-triplet Higgs
$M_{H_C}$ gets larger as the SUSY breaking scale $M_S$ is taken to be
higher. This originates from the mass difference among the components of
the fundamental Higgs multiplets, {\it i.e.}, the triplet-Higgs,
higgsinos, heavy Higgs bosons, and the lightest Higgs boson. Therefore,
the behavior of $M_{H_C}$ with respect to the SUSY breaking scale is
universal in a sense. Further, $M_{H_C}$ depends only on the ratio of
$M_2$ and $M_3$. $M_X^2M_\Sigma$ is, on the other hand, independent of
the SUSY braking scale $M_S$ while dependent on the scale of the
gauginos, not their ratio. This is because the right-hand side of
Eq.~\eqref{mgut} results from the mass difference in the gauge vector
multiplets and the adjoint Higgs multiplet, a part of which is included
as the longitudinal mode of the gauge multiplets. It is also found that
$M_X^2M_\Sigma$ decreases when the gaugino masses are taken to be large
values. This is owing to the opposite sign of the contribution of 
gauge fields to the gauge beta functions to those of 
matter fields. This feature is,
therefore, again model-independent. In the subsequent section, we carry
out a similar analysis using the two-loop RGEs.

%%%%%%%%%%%%%%%%%%%%%%%%%%%%%%%%%%%%
\section{Results}
\label{results}
%%%%%%%%%%%%%%%%%%%%%%%%%%%%%%%%%%%

%%%%%%%%%%%%%%%%%%%%%%%%%%%%%%FIGURE%%%%%%%%%%%%%%%%%%%%%%%%%%%%
\begin{figure}
  \centering
\includegraphics[height=7.5cm]{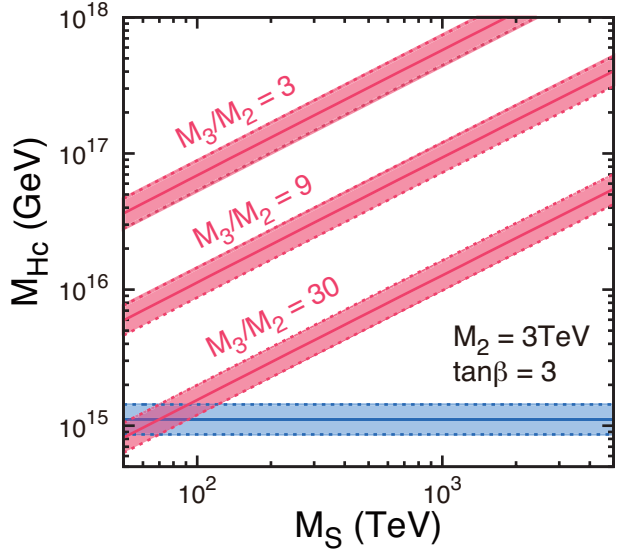}
\caption{Predicted color-triplet Higgs mass $M_{H_C}$ as
 functions of the SUSY breaking scale $M_S$ (pink lines). Here, wino
 mass $M_2$ is fixed to be 3~TeV and $\tan\beta=3$. Gluino-wino mass
 ratio, $M_3/M_2$, is set to be $M_3/M_2=3,~9,$ and $30$ from top to
 bottom, respectively. Theoretical errors coming from the strong
 coupling constant $\alpha_s(m_Z)=0.1184(7)$ \cite{Beringer:1900zz} are
 also shown. Horizontal blue line shows a result in the case of low-energy
 SUSY ($M_S=1$~TeV, $M_2=200$~GeV, and $M_3/M_2=3.5$).}
\label{tanb3mhc}
 \end{figure}
%%%%%%%%%%%%%%%%%%%%%%%%%%%%%%%%%%%%%%%%%%%%%%%%%%%%%%%%%%%%%%%%

%%%%%%%%%%%%%%%%%%%%%%%%%%%%%%FIGURE%%%%%%%%%%%%%%%%%%%%%%%%%%%%
\begin{figure}
  \centering
\includegraphics[height=7.5cm]{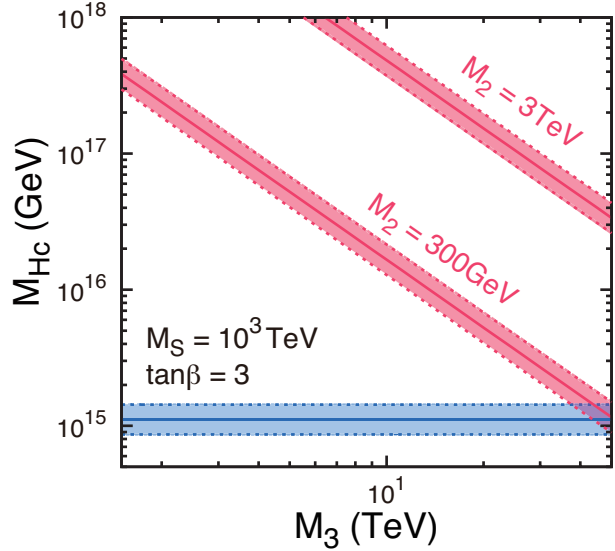}
\caption{Color-triplet Higgs mass $M_{H_C}$ as
 functions of gluino mass $M_3$ (pink lines). Here, $\tan\beta=3$ and
 $M_S=10^3$~TeV. Upper and lower lines correspond to $M_2=3$~TeV and
 300~GeV, respectively. Error bars indicate the input error of the strong
 coupling constant $\alpha_s(m_Z)=0.1184(7)$ \cite{Beringer:1900zz}.
Horizontal blue line shows a result in the case of low-energy
 SUSY ($M_S=1$~TeV, $M_2=200$~GeV, and $M_3/M_2=3.5$).}
\label{m3mhc}
 \end{figure}
%%%%%%%%%%%%%%%%%%%%%%%%%%%%%%%%%%%%%%%%%%%%%%%%%%%%%%%%%%%%%%%%

Now we present some results for the RGE analysis which we discuss in the
previous section. As noted above, the running of the gauge couplings is
computed at two-loop level and the threshold corrections are taken into
account at one-loop level. The masses of sfermions, heavy Higgs bosons,
and higgsinos are taken to be $M_S$ for brevity. Gaugino masses are
assumed to be lighter than $M_S$ by an one-loop factor.

First, we consider the color-triplet Higgs mass $M_{H_C}$. In
Fig.~\ref{tanb3mhc}, we plot the dependence of $M_{H_C}$ on the SUSY
breaking scale $M_S$ in the pink lines. Here, the wino mass $M_2$ is
fixed to be 3~TeV, which is favored from the thermal relic abundance,
and 
$\tan\beta=3$. The ratio of the gluino and wino masses, $M_3/M_2$, is
set to be $M_3/M_2=3,~9,$ and $30$ from top to bottom,
respectively. Further, we show the error of the calculation coming from
that of the strong coupling constant $\alpha_s(m_Z)=0.1184(7)$
\cite{Beringer:1900zz}. The horizontal blue line shows a result in
the case of low-energy SUSY ($M_S=1$~TeV, $M_2=200$~GeV, and
$M_3/M_2=3.5$) as a reference. In this case, we have $8.6\times
10^{14}\le M_{H_C}\le 1.4\times 10^{15}~{\rm GeV}$. This figure well
illustrates the feature read 
from the approximated expression given in Eq.~\eqref{mhc}; $M_{H_C}$
increases as the SUSY breaking scale grows while it decreases when the
ratio $M_3/M_2$ becomes large. To see the latter feature more clearly,
we show its dependence on the gluino mass $M_3$. Again, we set
$\tan\beta=3$, and the SUSY breaking mass is fixed to be
$M_S=10^3$~TeV. The upper and lower lines correspond to $M_2=3$~TeV and
300~GeV, respectively. These two figures show that $M_{H_C}$ is strongly
dependent on $M_S$ and $M_3/M_2$. Therefore, to predict the mass with
high accuracy, precise determination of the masses of gauginos as well
as the SUSY breaking scale is inevitable. Any way, in the high-scale
SUSY scenario it is found to be
possible for the mass of the color-triplet $M_{H_C}$ to be around $\sim
2\times 10^{16}$~GeV, which is expected by the gauge coupling
unification.

%%%%%%%%%%%%%%%%%%%%%%%%%%%%%%FIGURE%%%%%%%%%%%%%%%%%%%%%%%%%%%%
\begin{figure}
  \centering
\includegraphics[height=7.5cm]{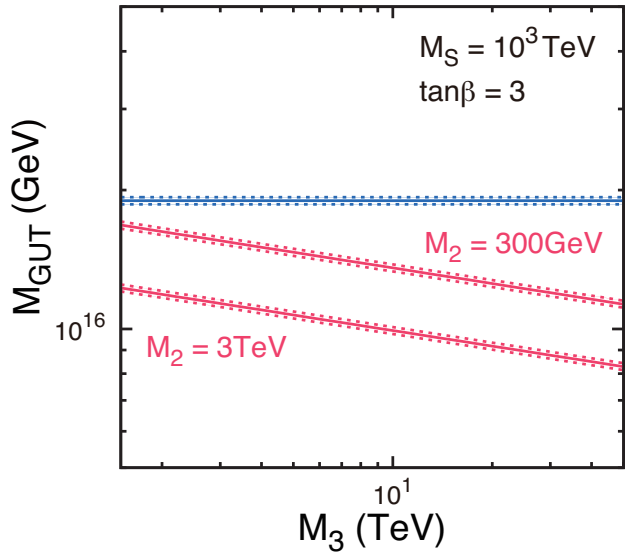}
\caption{GUT scale $M_{\rm GUT}\equiv (M_X^2 M_\Sigma)^{1/3}$ as
 functions of gluino mass $M_3$ (pink lines). Here, $\tan\beta=3$ and
 $M_S=10^3$~TeV. Upper and lower lines correspond to $M_2=300$~GeV and
 3~TeV, respectively. Error bars indicate the input error of the strong
 coupling constant $\alpha_s(m_Z)=0.1184(7)$ \cite{Beringer:1900zz}.
Horizontal blue line shows a result in the case of low-energy
 SUSY ($M_S=1$~TeV, $M_2=200$~GeV, and $M_3/M_2=3.5$).} 
\label{m3mgut}
 \end{figure}
%%%%%%%%%%%%%%%%%%%%%%%%%%%%%%%%%%%%%%%%%%%%%%%%%%%%%%%%%%%%%%%%

Next, we discuss constraints on $M_X^2 M_\Sigma$ derived from the relation
\eqref{mgut}. From now on we define $M_{\rm GUT}\equiv (M_X^2
M_\Sigma)^{1/3}$ and refer to it as the GUT scale. The equation
\eqref{mgut} tells us that the GUT scale depends on only the gaugino
masses at the leading order, so we express $M_{\rm GUT}$ as functions of
the gaugino masses. In Fig.~\ref{m3mgut} we plot it as functions of
gluino mass. Here again we fix $\tan\beta=3$ and $M_S=10^3$~TeV. The
upper and lower lines correspond to $M_2=300$~GeV and 3~TeV,
respectively. Again, the horizontal blue line
shows a result in the case of low-energy SUSY with $M_S=1$~TeV,
$M_2=200$~GeV, and $M_3/M_2=3.5$, which gives $M_{\rm GUT}\simeq
1.9\times 10^{16}$~GeV. The error bars indicate the input error of the strong
coupling constant $\alpha_s(m_Z)=0.1184(7)$ \cite{Beringer:1900zz},
though the effect is negligible. We see that the GUT scale also has little
dependence on the gaugino masses. In that sense, the prediction is robust
compared with that for $M_{H_C}$. However, as discussed in the previous
section, 
the GUT scale $M_{\rm GUT}$ gets lower when the gaugino masses become
larger ($M_{\rm GUT}\propto (M_3 M_2)^{-1/9}$).
This feature is quite interesting when one considers the proton
decay via the $X$-boson exchange processes. Although the change in
$M_{\rm GUT}$ is small, it might be significant since the proton decay
lifetime scales as $\propto M_X^4$.  For instance, when $M_X\simeq
0.8\times 10^{16}$~GeV, the proton lifetime via the $X$-boson exchange
reduces to around $5\times 10^{34}$ years \cite{Hisano:2012wq}, which is
slightly above the current experimental bound, $\tau(p\to e^+ \pi^0)>
1.29\times 10^{34}$~yrs \cite{Nishino:2012rv}.

Finally, we briefly comment on the $\tan\beta$ dependence of the
results. Although the one-loop computation is not related with
$\tan\beta$, the two-loop results might be affected through the running
of the Yukawa couplings. We have found, however, that the effects on the
results are negligible.

%%%%%%%%%%%%%%%%%%%%%%%%%%%%%%%%%%%%%%
\section{Conclusions and Discussion}
\label{conclusion}
%%%%%%%%%%%%%%%%%%%%%%%%%%%%%%%%%%%%

In this Letter, we have presented constraints on the masses of the GUT
scale particles from the gauge coupling unification in the case of the
high-scale SUSY scenario. To that 
end, we have used the two-loop RGEs for the gauge couplings with
one-loop threshold corrections considered. As a result, the mass of the
color-triplet Higgs multiplets $M_{H_C}$ turns out to be considerably large
compared with previous results in the traditional low-energy SUSY
scenario, while the GUT scale is found to be slightly lower. These are
generic features resulting from the mass difference among the components
of the same supermultiplet of the SU(5) gauge group. 
Interestingly, all of the superheavy particles might be around
$10^{16}$~GeV in the high-scale SUSY models.

The mass spectrum of the GUT scale particles predicted here stimulates
us to reconsider the proton decay in the case of high-scale
supersymmetry. As mentioned to above, the relatively low GUT scale
enhances the proton decay rate via the $X$-boson exchanging process. If
the enhancement is strong enough, the proton decay in the $p\to e^+
\pi^0$ channel might be searched in future experiments. Furthermore,
the experiments may also reach the proton decay through the
color-triplet Higgs exchange. Since the decay process predicts too
short lifetime \cite{Murayama:2001ur}, some suppression mechanism for the
process has been assumed. Nevertheless, with $M_{H_C}$ larger than
those considered in the previous literature and sfermions much heavier
than the electroweak scale, this process might evade the current
experimental bound without any mechanism of limiting the color-triplet
Higgs exchanging process. In such a case, it is possible for the $p\to
K^+\bar{\nu}$ mode, which is the main decay mode in the case of the
color-triplet Higgs exchange, to be searched in near future. A detailed
analysis of this decay process will be given elsewhere \cite{HKKN}.

Note that, in proposed models where the color-triplet Higgs exchange
is suppressed by some mechanism, large threshold corrections to
the gauge coupling constants at the GUT scale tend to appear. One of
the examples is introduction of the Peccei-Quinn symmetry
\cite{Peccei:1977hh}. It was found that when the threshold corrections
at the GUT scale are small, the suppression mechanism does not work
\cite{Hisano:1992ne,Hisano:1994fn}. In the SUSY SU(5) GUTs in higher
dimensional space the U(1)$_R$ symmetry forbids the dimension-five proton
decay, though the Kaluza-Klein particles generate large threshold
corrections to the gauge coupling constants \cite{Hall:2001pg}. The
high-scale supersymmetry would be another solution for the proton
decay problem, in which such large threshold corrections are not
required.

%%%%%%%%%%%%%%%%%%%%%%%%%%%%%%%%%%%%
\section*{Acknowledgments}
%%%%%%%%%%%%%%%%%%%%%%%%%%%%%%%%%%%%

The work of N.N. is supported by Research Fellowships of the Japan Society
for the Promotion of Science for Young Scientists. The work of
J.H. is supported by Grant-in-Aid for Scientific research from the
Ministry of Education, Science, Sports, and Culture (MEXT), Japan,
No. 20244037, No. 20540252, No. 22244021 and No. 23104011, and also by
World Premier International Research Center Initiative (WPI
Initiative), MEXT, Japan.

%%%%%%%%%%%%%%%%%%%%%%%%%%%%%%%%%%%%
{}
%%%%%%%%%%%%%%%%%%%%%%%%%%%%%%%%%%%%

\end{document}